\documentclass{article}
\usepackage{amsmath,amssymb,amsthm}
\usepackage{graphicx}
\usepackage{subcaption}
\usepackage{verbatim}
\usepackage{hyperref}
\usepackage{bm}
\usepackage{color}
\usepackage{amsmath}
\usepackage{amsfonts}
\usepackage{amssymb}
\usepackage[numbers]{natbib}

\begin{document}

\title{Surface Anchoring Energy of Cholesteric Liquid Crystals}
\author{Tianyi Guo \\
%EndAName
\emph{Advanced Materials and Liquid Crystal Institute, Kent State
University, OH, USA}\\
Xiaoyu Zheng \\
\emph{Department of Mathematical Sciences, Kent State University, OH, USA}\\
Peter Palffy-Muhoray \\
\emph{Advanced Materials and Liquid Crystal Institute, Kent State
University, OH, USA} }
\date{}
\maketitle

\begin{abstract}
In this paper, we propose a suitable surface energy expression for
cholesteric liquid crystals. We show that there exists a symmetry allowed
term for chiral nematics that doesn't appear in the traditional
Rapini-Papoular surface energy form. We discuss some consequences of this
new surface anchoring term.
\end{abstract}


\section{Introduction}

In this paper, we consider the surface energy density of cholesteric liquid
crystals(CLCs). Symmetry arguments have been successfully used to predict
novel phenomena in liquid crystals \cite{Mayer}. Using symmetry arguments,
we show that in addition to the standard Rapini-Papoular form \cite{Rapini},
a new additional term exists, which may change the behavior of cholesteric
samples.

Symmetry arguments rely on tensor representations which allow
coordinate-system independent covariant expressions to be formed. One
identifies the relevant tensor quantities and then inquires how these can be
combined to form an expression with the right tensor properties. All
physical quantities are tensors of different rank; scalars are rank zero,
vectors are rank one, and so on. Tensors are either `proper' or `pseudo-';
components of proper tensors of even rank do not change sign under parity
(centro-inversion of the coordinate system) but those of odd rank do, while
components of pseudotensors of odd rank do not change sign under parity, but
those of even rank do. Basic physical quantities (length, mass, time,
temperature, electric current) are proper; other quantities derived from
these may be proper, such as energy, force, electric field, and dielectric
permittivity, or pseudo, such as torque, angular momentum, and magnetic
field.

Symmetry arguments are based on two key points. If the quantity of interest
with the right property (proper or pseudo) cannot be constructed from the
set of relevant quantities, then the existence of the quantity of interest
arising from the relevant quantities is forbidden by symmetry, and the
corresponding process cannot occur. If the quantity of interest with the
right property (proper of pseudo) can be constructed from the set of
relevant quantities, then the quantity of interest is allowed by symmetry,
and will consequently occur in general. \footnote{%
Totalitariansm has been adopted as a principle of physics; according to
Gell-Mann``Everything not forbidden is compulsory." \cite{Gell-Mann}}

The paper is organized as follows. In Section 2, a new term in the surface
anchoring energy of cholesteric liquid crystals is proposed using a symmetry
argument. We summarize and conclude the paper in Section 3.

\section{Surface anchoring energy}

Liquid crystals are characterized by orientational order. Orientational
order can be modeled in a variety of ways; here we begin by using the order
parameter tensor formalism. The orientation of liquid crystal molecules is
described by a unit vector $\mathbf{\hat{l}}$, and the order parameter tensor $%
Q_{\alpha \beta }$ is defined by%
\begin{equation}
Q_{\alpha \beta }=\left\langle \frac{1}{2}(3l_{\alpha }l_{\beta }-\delta
_{\alpha \beta })\right\rangle ,
\end{equation}%
where $\left\langle {}\right\rangle $ denote ensemble average. The order
parameter tensor is symmetric and traceless, it may be uniaxial with two
identical eigenvalues, or biaxial, where all three eigenvalues are different.

Since liquid crystals are soft, for large samples, bulk orientation is
dominated by surface interactions. Surface anchoring is therefore of
paramount importance both in fundamental studies and also in device
applications.

We now use symmetry arguments to construct the allowed surface anchoring
energy of nematic liquid crystals. In order to identify the relevant
quantities, we consider the interaction of nematic molecules with a
substrate. There are both steric and attractive interaction with the
substrate, as with other liquid crystal molecules in the bulk. For this
reason, it has been proposed that the substrate may be regarded as
consisting of `frozen' liquid crystal molecules, with some given order
parameter tensor $Q_{\alpha \beta }^{0}$ \cite{Durand} . Using this, we
identify the relevant quantities to form the surface anchoring energy
density as the liquid crystal order parameter $Q_{\alpha \beta }$, the
surface normal unit vector $N_{\alpha }$, and the preferred anchoring order
parameter at the substrate $Q_{\alpha \beta }^{0}$.

The order parameters can be written in different forms, choice is a matter
of convenience. We note that the liquid crystal order parameter can be
written as%
\begin{equation}
Q_{\alpha \beta }=\lambda _{1}n_{\alpha }n_{\beta }+\lambda _{2}m_{\alpha
}m_{\beta },
\end{equation}%
or%
\begin{equation}
\mathbf{Q}=\lambda _{1}\mathbf{\hat{n}\hat{n}+}\lambda _{2}\mathbf{\hat{m}%
\hat{m}},
\end{equation}%
where $\mathbf{\hat{n}\hat{n}}$ is a dyad, the tensor product of the nematic
director $\mathbf{\hat{n}}$ with itself. If the system is biaxial, director
defining the direction of the minor axis is $\mathbf{\hat{m}}$. Similarly,
we write for the preferred anchoring order parameter at the substrate $%
Q_{\alpha \beta }^{0}$ as
\begin{equation}
\mathbf{Q}^{0}=\lambda _{1}^{0}\mathbf{\hat{n}}^{0}\mathbf{\hat{n}}^{0}%
\mathbf{+}\lambda _{2}^{0}\mathbf{\hat{m}}^{0}\mathbf{\hat{m}}^{0},
\end{equation}%
and we have the surface normal $\mathbf{\hat{N}}$.

\subsection{Achiral nematics}

Here we consider achiral nematics, where both the phase and the particles
have inversion symmetry. We next consider forming the surface anchoring
energy density, a proper scalar, from the relevant quantities $Q_{\alpha
\beta }$, $N_{\alpha }$ and $Q_{\alpha \beta }^{0}$. We can construct, to
lowest order in $Q_{\alpha \beta }$, three non-zero terms. We list all of
these in the expression for the achiral anchoring energy density 
\begin{equation}\label{eq5}
\mathcal{E}_{a}\mathcal{=}\frac{1}{2}W_{1}Q_{\alpha \beta }Q_{\alpha \beta
}^{0}+\frac{1}{2}W_{2}Q_{\alpha \beta }N_{\alpha }N_{\beta }+\frac{1}{2}%
W_{3}Q_{\alpha \beta }N_{\alpha }Q_{\beta \gamma }^{0}N_{\gamma }.
\end{equation}%
This can be written more simply as 
\begin{equation}
\mathcal{E}_{a}\mathcal{=}\frac{1}{2}W_{1}\mathbf{Q:Q}^{0}+\frac{1}{2}W_{2}%
\mathbf{Q}:\mathbf{\hat{N}\hat{N}}+\frac{1}{2}W_{3}(\mathbf{Q\cdot \hat{N}%
)\cdot (Q}^{0}\mathbf{\cdot \hat{N})},
\end{equation}%
and if the system is uniaxial, we recover 
\begin{equation}
\mathcal{E}_{a}\mathcal{=}\frac{1}{2}W_{1}^{\prime }\mathbf{(\hat{n}\cdot 
\hat{n}}^{0}\mathbf{)}^{2}+\frac{1}{2}W_{2}^{\prime }\mathbf{(\hat{n}\cdot 
\hat{N})}^{2}+\frac{1}{2}W_{3}^{\prime }(\mathbf{\hat{n}\cdot \hat{N})(\hat{n%
}}^{0}\mathbf{\cdot \hat{N})}(\mathbf{\hat{n}\cdot \hat{n}}^{0}),
\end{equation}%
which has the usual Rapini-Papoular form if $\mathbf{\hat{n}}^{0}$ is in the
plane of the substrate (that is, where $\mathbf{\hat{n}}^{0}\mathbf{\cdot 
\hat{N}}=0$), while the third describes anchoring energy with pre-tilt.

If the rubbing direction is in the plane of the surface, the director will
align either along the surface normal $\mathbf{\hat{N}}$ or the rubbing
direction $\mathbf{\hat{n}}^{0}$. If rubbing is not in the plane, the third
term allows the director to orient in other directions.

\subsection{Chiral nematics}

Now we turn to the central point of our paper, surface anchoring of chiral
nematic or cholesteric liquid crystals. The novel aspect of cholesterics is
broken inversion symmetry due to the chirality of the constituent molecules.
Since chiral objects lack inversion symmetry, it is possible to associate three
non-collinear proper vectors with chiral systems. The three proper vectors $%
\mathbf{A}$, $\mathbf{B}$ and $\mathbf{C}$ can be combined to form the
pseudoscalar $\mathbf{A\cdot (B\times C)}$. Since the converse is also true,
(a handed vector triad can be constructed if a pseudoscalar exists),
pseudoscalars are indicators of chirality. In the case of cholesteric liquid
crystals, a pseudoscalar $q_{0}$ is often used to represents the intrinsic
twist in the orientation of molecules, with $q_{0}>0$ for right-handed
chirality, and $q_{0}<0$ for left-handed chirality. The wavelength of the
periodicity of the system is the pitch $p=2\pi /|q_{0}|$. With a
pseudoscalar $q_{0}$ in the system, one can construct one additional proper
scalar term, arising from chirality, in the surface anchoring energy
density. The new term, in addition to the ones in Eq.\ (\ref{eq5}), is%
\begin{equation}
\mathcal{E}_{c}\mathcal{=}\frac{1}{2}q_0 W_{c}\varepsilon _{\alpha \beta
\gamma }Q_{\mu \beta }N_{\gamma }Q_{\mu \alpha }^{0} ,  \label{main}
\end{equation}%
where $\varepsilon _{\alpha \beta \gamma }$ is the Levi-Civita antisymmetric
tensor. The energy $\mathcal{E}_{c}$ is a proper scalar, allowed by
chirality and the presence of the pseduoscalar $q_{0}$. Explicitly, in terms
of the eigenvectors, it becomes%
\begin{eqnarray}
\mathcal{E}_{c} &\mathcal{=}&\frac{1}{2}q_{0}\{W_{c1}\mathbf{\hat{n}}%
^{0}\cdot (\mathbf{\hat{N}\times \hat{n}})(\mathbf{\hat{n}\cdot \hat{n}}%
^{0})+W_{c2}\mathbf{\hat{m}}^{0}\cdot (\mathbf{\hat{N}\times \hat{n}})(%
\mathbf{\hat{n}\cdot \hat{m}}^{0})  \notag \\
&&+W_{c3}\mathbf{\hat{n}}^{0}\cdot (\mathbf{\hat{N}\times \hat{m}})(\mathbf{%
\hat{m}\cdot \hat{n}}^{0})+W_{c3}\mathbf{\hat{m}}^{0}\cdot (\mathbf{\hat{N}%
\times \hat{m}})(\mathbf{\hat{m}\cdot \hat{m}}^{0})\}.
\end{eqnarray}%
We note that the energy $\mathcal{E}_{c}$, a proper scalar, does not change
sign under centroinversion.

\begin{figure}[th]
\centering
\includegraphics[width=.5\linewidth]{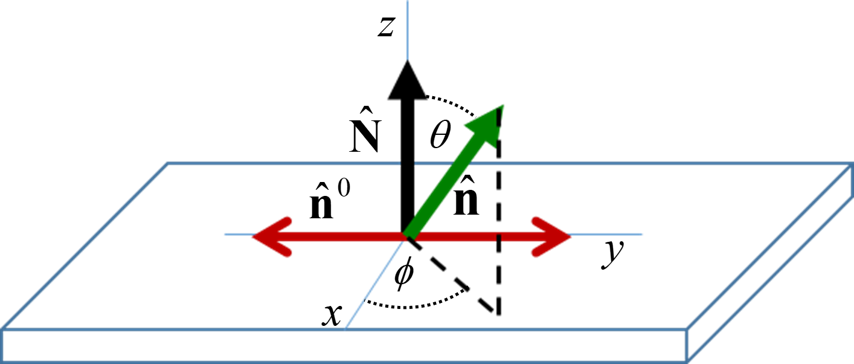}
\caption{A schematic showing the cholesteric liquid crystals and a
substrate. Here $\mathbf{\hat{N}}$ represents outward surface normal, $%
\mathbf{\hat{n}}^0$ is the rubbing direction, and $\mathbf{\hat{n}}$ is the
chiral nematic director.}
\label{fig_schematic}
\end{figure}

In the uniaxial case, the above expression simplifies to%
\begin{equation}
\mathcal{E}_{c}\mathcal{=}\frac{1}{2}q_{0}W_{c1}\mathbf{\hat{n}}^{0}\cdot (%
\mathbf{\hat{N}\times \hat{n}})(\mathbf{\hat{n}\cdot \hat{n}}^{0}),
\end{equation}%
which clearly shows the contribution of chirality to the anchoring energy. The geometry
is shown in Fig.\,\ref{fig_schematic} for the case when the rubbing
direction $\mathbf{\hat{n}}^{0}$ is in the plane of the substrate.

In terms of the angles, the new term due to chirality becomes

\begin{equation}
\mathcal{E}_{c}\mathcal{=}\frac{1}{2}q_{0}W_{c1}\sin ^{2}\theta \sin 2\phi,
\end{equation}
which is minimized when the director is at $45^o$ or $135^{o}$ from the $x-$axis, the 
direction depends on the sign of the coefficient. When the rubbing direction $\mathbf{\hat{n}}^{0}$ is in the plane
of the substrate, the total energy is given by 
\begin{equation}
\mathcal{E}_{t}\mathcal{=}\frac{1}{2}W_{1}^{\prime }\mathbf{(\hat{n}\cdot 
\hat{n}}^{0}\mathbf{)}^{2}+\frac{1}{2}W_{2}^{\prime }\mathbf{(\hat{n}\cdot 
\hat{N})}^{2}+\frac{1}{2}q_{0}W_{c1}\mathbf{\hat{n}}^{0}\cdot (\mathbf{\hat{N%
}\times \hat{n}})(\mathbf{\hat{n}\cdot \hat{n}}^{0}).
\end{equation}%
The minimum energy is obtained when the director is either along the normal $%
\mathbf{\hat{N}}$, or perpendicular to it, along $\mathbf{\hat{n}}^{0}$. It the latter case,
the equilibrium orientation of the director will differ from the rubbing
direction. The magnitude of the deviation depends on the relative strengths
of the surface interactions, and direction of the deviation is the opposite
for right - and left-handed systems.

Dimensional analysis suggests that the ratio of $W_{c1}/W_{1}$ is small, of
the order of the molecular length $l_{0}$ to the pitch $p$. This is only a
conjecture, however, to be verified by experiment. Many chiral nematic
systems consist of achiral nematic molecules mixed with chiral dopants. It
is expected that in such systems, due the presence of the new anchoring
term, the concentration of chiral dopants may be different at the surface
then at the bulk. Verifying this may be another experimental verification of
our model.

We note that our strategy used above can be adapted to other systems. For
example, in a bulk chiral liquid crystal elastomer with strain, one can
construct an energy term similar to that in Eq.\,(\ref{main}), where the
strain tensor plays the role of $\mathbf{Q}^{0}$ and, say, the electric
field the role of $\mathbf{\hat{N}}$. Such a term predicts a linear
mechanical response of cholesteric elastomers to an applied electric field.

\section{Summary}

In this paper, we propose a general form of surface anchoring energy of
chiral liquid crystals. Explicitly, we argue that in addition to existing
terms in the literature, there exists a new symmetry allowed term in the
surface anchoring energy density. Chirality leads to the deviation of the
director at equilibrium from the rubbing direction. Our results may also be
useful for novel applications of chirality-induced phenomena \cite{Yossi},
and our strategy may be useful in predicting novel phenomena in chiral
systems.

\paragraph{Acknowledgements}

Subsequent to this work, we became aware of a similar result by E.C.
Gartland \cite{Gartland}.  We acknowledge support for this work by the Office of Naval Research
through the MURI on Photomechanical Material Systems (ONR N00014-18-1-2624).

\end{document}